\documentclass{aa}  
\usepackage{graphicx}
\usepackage{epsfig}
\usepackage{natbib}
\usepackage{lscape}
\usepackage{rotating}
\usepackage{tabularx}
\usepackage{subfig}
\bibpunct{(}{)}{;}{a}{}{,}
\usepackage{txfonts}
\begin{document} 

\titlerunning{Evidence for Rapid Variability in SN 2014J}
\authorrunning{Bonanos and Boumis}

   \title{Evidence for Rapid Variability in the Optical Light Curve \\of
     the Type Ia SN~2014J\thanks{Based on observations made with the
       2.3m Aristarchos telescope, Helmos Observatory, Greece, which is
       operated by the Institute for Astronomy, Astrophysics, Space
       Applications and Remote Sensing, National Observatory of Athens,
       Greece.}\fnmsep\thanks{Table 2 is only available in electronic
       form via http://www.edpsciences.org}}

   \author{A. Z. Bonanos
          \and
          P. Boumis}
   \institute{IAASARS, National Observatory of Athens, GR-15236 Penteli,
     Greece\\ \email{bonanos@astro.noa.gr} }

   \date{Received November 26, 2014; accepted October 25, 2015}

\abstract{We present results of high-cadence monitoring of the optical
  light curve of the nearby, Type Ia SN~2014J in M82 using the 2.3m
  Aristarchos telescope. $B$ and $V-$band photometry on days 15--18
  after t$_{max}(B)$, obtained with a cadence of 2 min per band, reveals
  evidence for rapid variability at the 0.02--0.05 mag level on
  timescales of 15--60 min on all four nights, taking the red noise
  estimation at face value. The decline slope was measured to be steeper
  in the $B-$band than in $V-$band, and to steadily decrease in both
  bands from 0.15 mag day$^{-1}$ (night 1) to 0.04 mag day$^{-1}$ (night
  4) in $V$ and from 0.19 mag day$^{-1}$ (night 1) to 0.06 mag
  day$^{-1}$ (night 4) in $B$, corresponding to the onset of the
  secondary maximum. We propose that rapid variability could be due to
  one or a combination of the following scenarios: the clumpiness of the
  ejecta, their interaction with circumstellar material, the asymmetry
  of the explosion, or the mechanism causing the secondary maximum in
  the near-infrared light curve. We encourage the community to undertake
  high-cadence monitoring of future, nearby and bright supernovae to
  investigate the intraday behavior of their light curves.}
% 5 {} token are mandatory
 

   \maketitle
%
%________________________________________________________________

\section{Introduction}

Nearby supernovae offer the opportunity to explore the short-timescale
and low-amplitude variability properties of their light curves. Even
though several nearby, bright
%($V_{max}<13$~mag)
supernovae\footnote{http://www.rochesterastronomy.org/snimages/snmag.html}
(SNe) have been discovered recently \citep[e.g.\ the Type Ia SN~2011fe
  and SN~2013aa, the Type IIP SN~2013ej;][respectively]{Nugent11,
  Waagen13, Richmond14}, no high-cadence variability search has been
undertaken so far to explore variability on timescales of minutes or
hours. Typical photometric monitoring of SNe consists of a single
observation per night or every few nights in several filters, therefore,
the intraday behavior of the light curves of SNe remains uncharted
territory.\footnote{While this paper was under review, \citet{Olling15}
  reported serendipitous, high-cadence monitoring (every 30 min) of
  three type Ia supernovae by the Kepler Mission.}

To rectify the situation, we performed high-cadence photometry of the
nearby Type Ia supernova (SN Ia) SN 2014J \citep[see e.g.][and
  references therein]{Foley14}, which was discovered in the galaxy M82
in January 2014 \citep{Fossey14}. Reaching $V_{max}=10.61\pm0.05$~mag
\citep{Foley14}, it was ideally suited for a variability search, which
we conducted in February 2014 with the 2.3m Aristarchos telescope. Just
before the submission of this work, \citet{Siverd15} reported relatively
high-cadence photometry of SN 2014J with the 4.2cm Kilodegree Extremely
Little Telescope North (KELT-N), thereby placing a $4.5\%$ (3$\sigma$)
upper limit on short timescale variations. This paper reports the
results of our monitoring survey: the observations and data reduction
are presented in Secion 2, the analysis in Section 3, the results in
Section 4, and the discussion and conclusions in Section 5.

%----------------------------------------------------------- 
   \begin{figure}[ht]
   \centering
   \includegraphics[width=\hsize, trim= 0 140 0 100, clip=true]{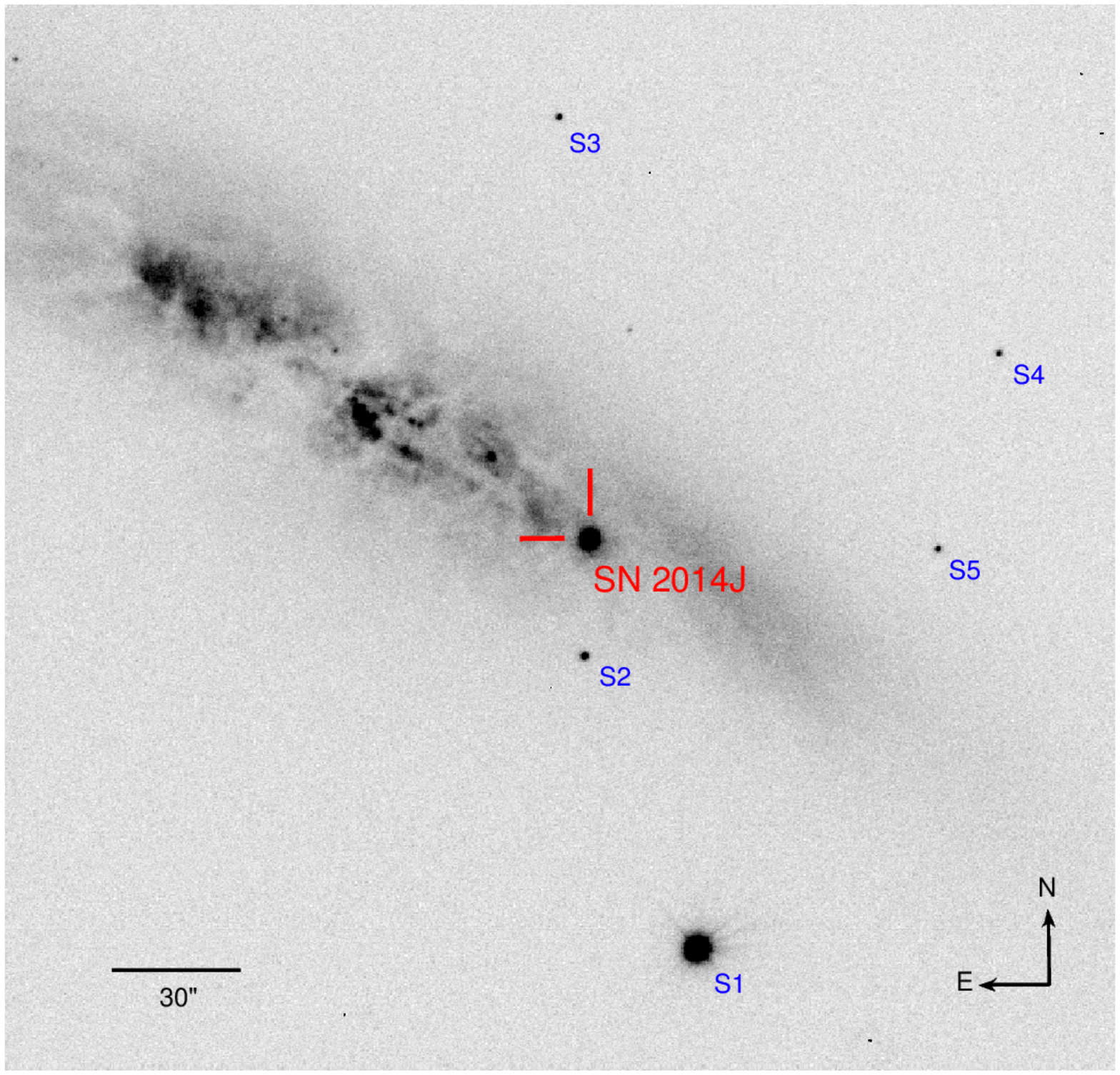}
      \caption{$V-$band image of SN 2014J in M82, obtained on 2014
        February 16 with the Aristarchos telescope. The positions of the
        5 comparison stars (S1--S5) used in the analysis are labelled.}
      \label{finder}
   \end{figure}
%
%______________________________________________________________

%----------------------------------------------------------- 
\begin{figure*}
%\begin{figure}
\centering
\includegraphics[angle=270, width=17cm]{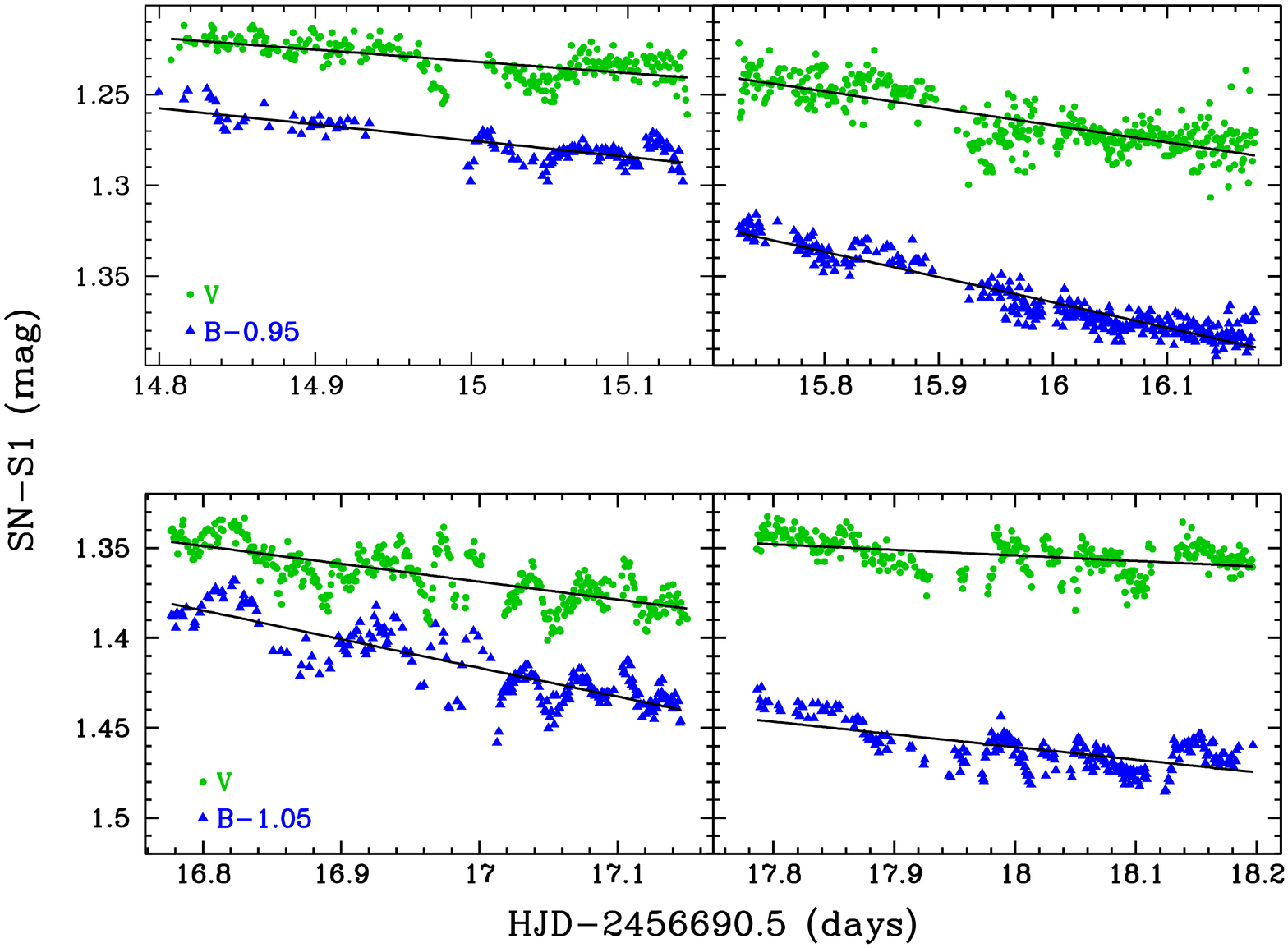}
\caption{Calibrated $V-$band (circles) and $B-$band (triangles)
  differential light curves of SN 2014J, compared with Star 1, as a
  function of the time in days measured since t$_{max}(B)=2456690.5$
  \citep[JD;][]{Foley14}. Each panel corresponds to one night of
  observations. The $B-$band curves are offset by 0.95 mag in the upper
  panels and 1.05 mag in the lower panels for clarity. The black lines
  represent least-square fits to the photometry.}
\label{lightcurve}
\end{figure*}
%\end{figure}

%
%______________________________________________________________

\section{Observations and Data Reduction}

SN 2014J was monitored for variability with the LN CCD camera on the
2.3m Aristarchos telescope for 4 consecutive nights: 2014 February
16--19 (nights 1--4 or N1--4), corresponding to days 14.8--18.2 after
t$_{max}(B)=2456690.5\pm0.2$ \citep[JD;][]{Foley14}. The
1024$\times$1024 SITeAB back-illuminated CCD has pixels that are
24~$\mu$m in size, which map to $0.28\arcsec$ pixel$^{-1}$ on the focal
plane, making each image $4.8\arcmin$ on the side. Figure~\ref{finder}
provides a finder chart, labelling the position of the SN and the
comparison stars (S1--S5) used in the analysis below.

SN 2014J was observed for about 8 hrs per night to obtain differential
photometry. Sequences of alternating $V-$band and $B-$band images were
obtained with 5 sec and 20 sec exposure times, respectively, yielding
400--500 images per band each night and a cadence of 2 min per
band. Small gaps in the data are mainly due to recentering performed in
between typical sequences of 100 images, as the absence of guiding
caused the target to drift during the sequences.

Initial reduction of the images was performed using standard routines in
the IRAF\footnote{IRAF is distributed by the National Optical Astronomy
  Observatory, which is operated by the Association of Universities for
  Research in Astronomy (AURA) under cooperative agreement with the
  National Science Foundation.} \textit{ccdproc} package (i.e.\ bias
level subtraction, flat field division). The images were aligned with
the IRAF \textit{imalign} task, while the seeing was calculated using
the IRAF task \textit{psfmeasure}. The statistical mode values for the
seeing in the $V-$band images each night were 1.2$\arcsec$\!,
1.4$\arcsec$\!, 1.2$\arcsec$\!, 1.2$\arcsec$\!, respectively, and
1.2$\arcsec$\!, 1.6$\arcsec$\!, 1.05$\arcsec$\!, 1.25$\arcsec$\! in the
$B-$band. Typical $\sigma$ values were 0.55$\arcsec$ for N2 and
$0.2-0.3\arcsec$ for the other nights. Note, on N1/N3, the S/N of the
maximum pixel value of S1 in $B$ (190/185) exceeded the mean S/N of the
$B-$flat (176/172).

\section{Analysis}

Aperture photometry was extracted from the supernova and the 5 brightest
comparison stars available in the field, using a 5 pixel or 1.4$\arcsec$
radius with the IRAF \textit{apphot} package and a 5 pixel wide sky
annulus 15 pixels from the source. Typical photometric errors are 1~mmag
for the SN and S1, 0.01~mag for S2, and 0.01--0.02 mag for S3--S5.

The instrumental magnitudes for the SN and S1 were transformed to the
standard system by computing zeropoint offsets and color terms, using
$B$ and $V$ magnitudes for SN~2014J on N1 from \citet{Foley14} and for
S1 from the APASS Catalogue\footnote{http://www.aavso.org/apass/}. The
resulting transformation equations are: 

%$V=v-1.320+0.056\cdot(B-V)$, and $B=b-2.609+0.005\cdot(B-V)$, where $b$
%and $v$ are the instrumental magnitudes.
\vspace{-0.3cm}
\begin{equation}
V=v-1.320+0.056\cdot(B-V)
\end{equation}
\begin{equation}
B=b-2.609+0.005\cdot(B-V)
\end{equation}
\noindent where $b$ and $v$ are the instrumental magnitudes.

%----------------------------------------------------------- 
   \begin{figure*}
   \centering
   \includegraphics[width=\hsize, trim= 0 40 0 0, clip=true]{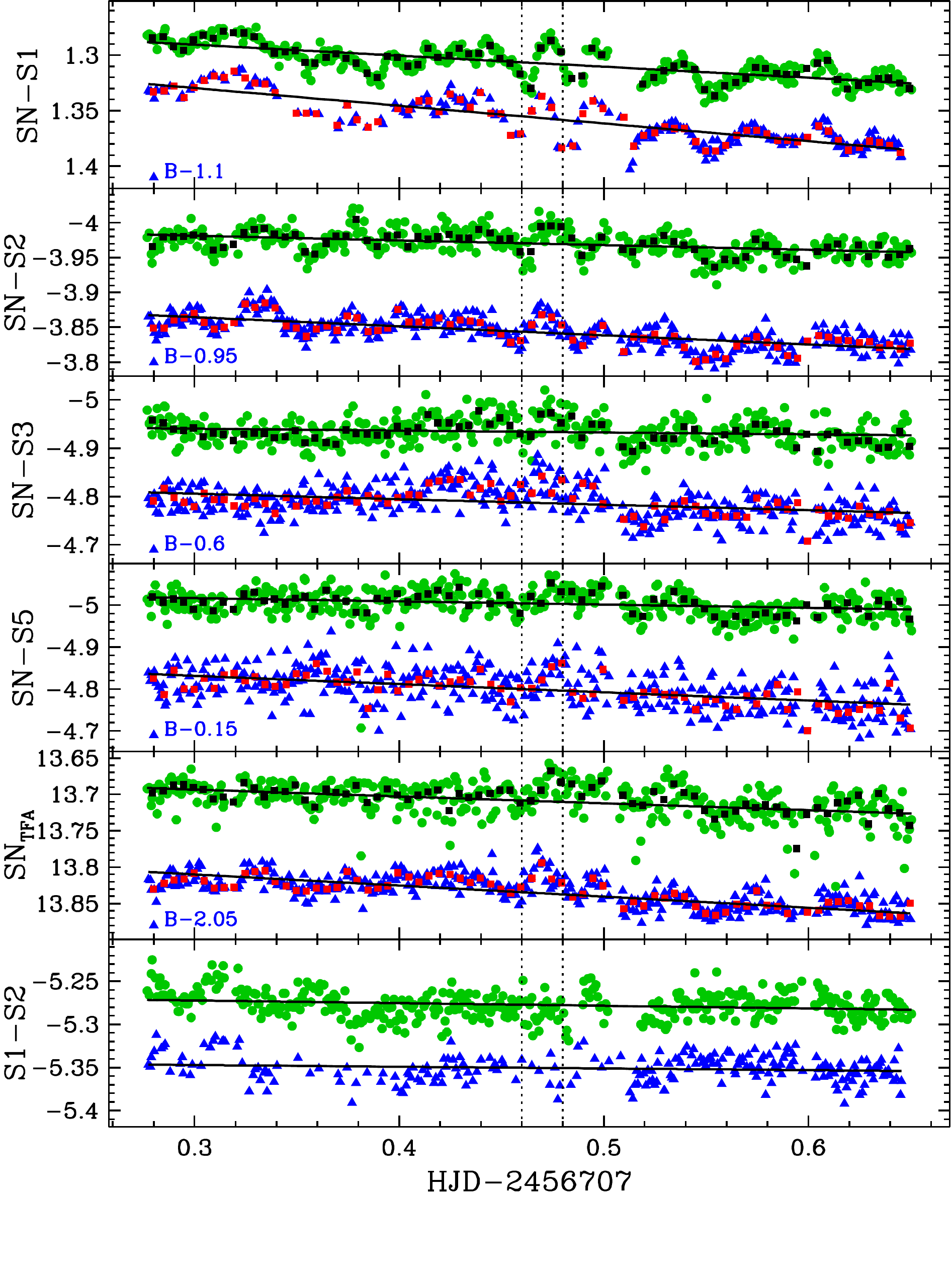}
      \caption{Instrumental $B-$band (blue triangles) and $V-$band
        (green circles) light curves of SN 2014J obtained on N3. The
        first four panels show the differential light curves of the SN
        with respect to S1--S3 and S5, respectively, followed by a panel
        showing the reconstructed light curves using the trend fitting
        algorithm (TFA) and the differential curve S1--S2. Solid lines
        correspond to least-square fits to the data, while black and red
        squares correspond to the binned $V$ and $B-$band light curve,
        respectively, using a bin size of 0.005 days. The $B-$band
        curves are offset for clarity by the amount indicated in each
        panel. Dotted lines indicate regions of significant
        variability.}
      \label{Fig3n3}
   \end{figure*}

\begin{figure*}
\includegraphics[width=\hsize, trim= 0 40 0 0, clip=true]{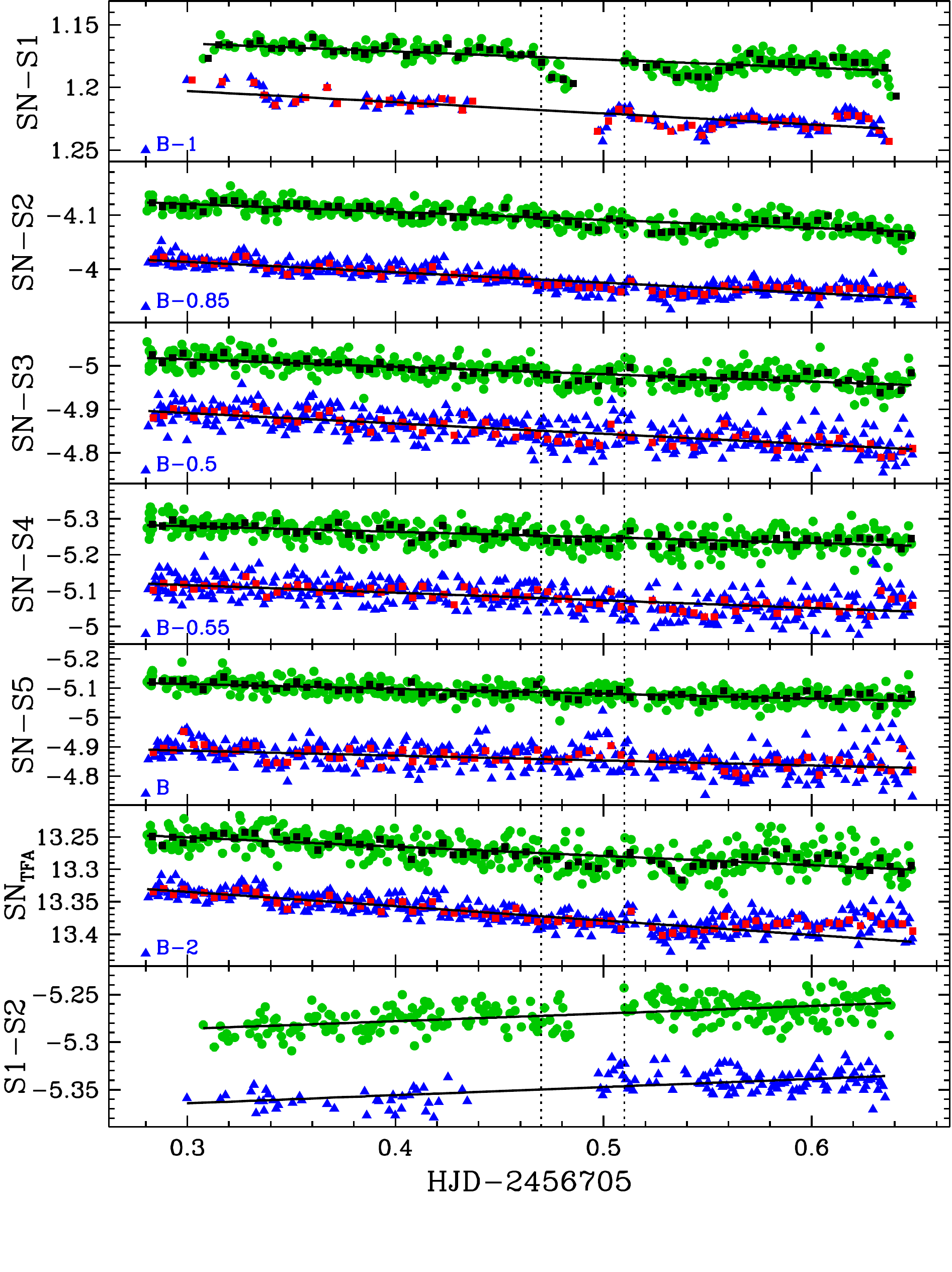}
\caption {Same as Figure~\ref{Fig3n3}, but for N1. On this night all 5
  comparison stars were available, therefore the first five panels show
  the differential light curves with respect to S1--S5, respectively,
  followed by a panel showing the reconstructed light curves using the
  trend fitting algorithm and the differential curve
  S1--S2.} \label{Fig3n1}
\end{figure*}

\begin{figure*}
\includegraphics[width=\hsize, trim= 0 40 0 0, clip=true]{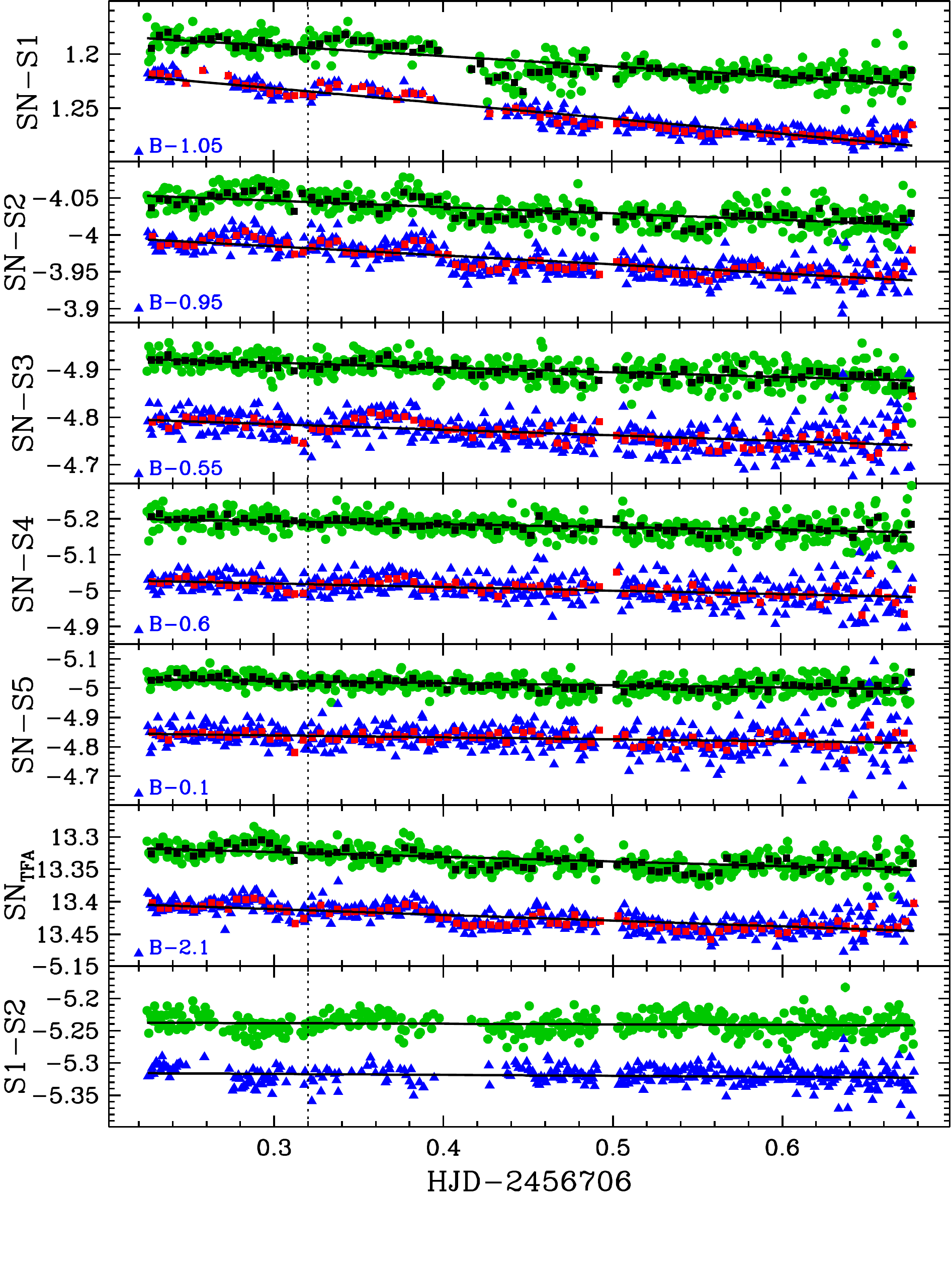}
\caption {Same as Figure~\ref{Fig3n1}, but for N2.} \label{Fig3n2}
\end{figure*}

\begin{figure*}
\includegraphics[width=\hsize, trim= 0 40 0 0, clip=true]{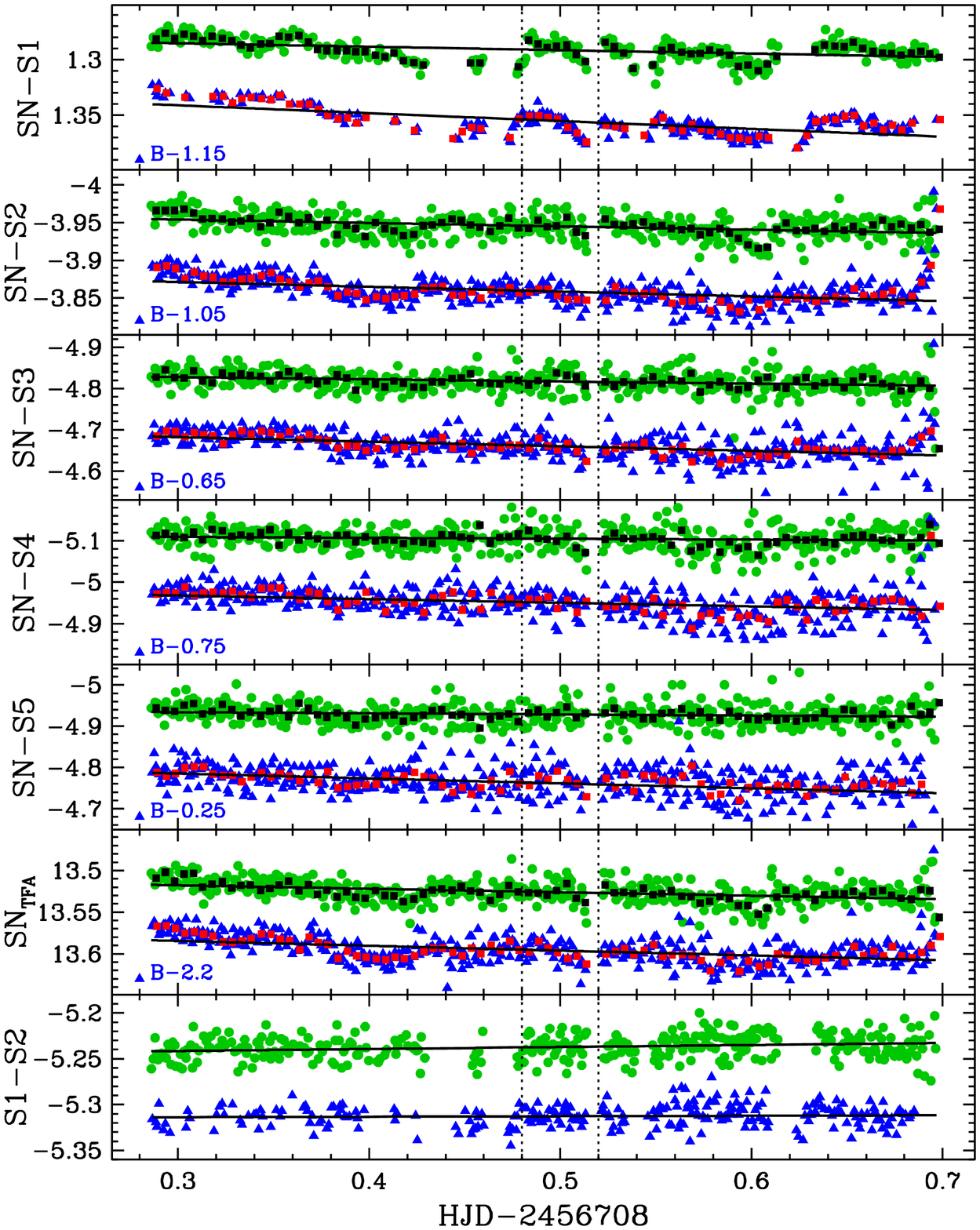}
\caption {Same as Figure~\ref{Fig3n1}, but for N4.} \label{Fig3n4}
\end{figure*}
%______________________________________________________________

Differential light curves were created by taking the difference between
the instrumental magnitude of the supernova and each of the 5 comparison
stars. S1 yields the smallest scatter, with an error in each point of
1.4 mmag. However, due to its brightness, it was saturated or exceeded
the non-linearity limit of the CCD ($\sim45000$ ADU) in the best seeing
$B-$band images, creating gaps in the differential curves with respect
to S1. Furthermore, the absence of guiding caused S1 to intercept a bad
column, creating spikes in the light curves in both bands, which were
selected visually and removed. Figure~\ref{lightcurve} presents the
differential, calibrated, light curves with respect to S1, which show
the expected decline of the SN and onset of the secondary maximum
discussed below.

Figures~\ref{Fig3n3}--\ref{Fig3n4} show the differential light curves
with respect to all comparison stars available each night. We used a
detrending algorithm to reconstruct the light curves, rule out intrinsic
variability in S1 (as no variability information for S1 is available in
the literature), and correct for systematic effects. We applied the
Trend Filtering Algorithm \citep[TFA;][]{Kovacs05}, as implemented in
the VARTOOLS package \citep{Hartman08}, using S2--S5 to reconstruct the
light curves of the SN and S1. Due to the fact that S2--S5 are fainter,
this method introduced noise at the level of $0.01-0.02$ mags to the
light curves of the brighter stars, however, we found that the
variability signal remains, although it is still affected by
systematics. The two lower panels of Figures~\ref{Fig3n3}--\ref{Fig3n4}
present the detrended light curves and the differential curve S1--S2,
respectively. The latter illustrates systematics present in the
photometry, which are discussed in the next Section. Features that are
present in all light curves (more clearly illustrated in the binned
curves) and that correspond to a featureless S1--S2 curve provide
evidence for rapid variability in the SN.

We next performed a test using artificial stars to assess the
variability signal seen in the differential light curves. We inserted
artificial stars to each of the $V-$band images taken on N3, which
displays the largest variability, using the appropriate PSF derived for
each image with the \textit{daophot} package in IRAF. We inserted 3
stars at the magnitude of the SN along the disk of the galaxy at
locations with both smoother and more complex galaxy background, 3 stars
at the magnitude of S2 at the same distance from the galaxy disk, and 3
stars at the magnitude and distance of S1. We then performed aperture
photometry, as described above, and constructed differential light
curves. These were found to be dominated by Poisson noise and not to
show any patterns resembling the varibility signal. In particular, the
standard deviation ($\sigma$) of the points for the 'artificial' SN in
all 3 positions tested were $4.5-4.8$ mmag, while we found $\sigma=1.6$
mmag for SN$-$S1, $\sigma=0.0115$ mag for SN$-$S2 and $\sigma=0.012$ mag
for S1$-$S2. The artificial star test thus provides additional support
for the variability signal originating in the SN.

\section{Results}

The densely sampled differential light curves shown in
Figures~\ref{lightcurve}--\ref{Fig3n4} provide a measure of the
intranight decline rate of SN~2014J. Table~\ref{table:decline} presents
the decline rates measured each night in $V$ and $B$ based on
least-square fits to the differential and detrended light curves shown
in Figures~\ref{Fig3n3}--\ref{Fig3n4}. The error bars represent the
root-mean-square (rms) error of the points to the fit. Note, the slopes
inferred for SN--S1 are identical for the calibrated and instrumental
light curves. All values are in agreement within errors, except for
those derived using SN--S1 on N1, due to the larger gaps in the
photometry on that night. Adopting the values derived from SN--S2,
SN~2014J declines with a rate $\alpha$ of $0.15\pm0.01$, $0.09\pm0.01$,
$0.07\pm0.02$, $0.04\pm0.01$ mag day$^{-1}$ in the $V-$band and
$0.19\pm0.01$, $0.12\pm0.01$, $0.13\pm0.02$, $0.06\pm0.02$ mag
day$^{-1}$ in the $B-$band\footnote{Note, no extinction or atmospheric
  corrections were applied.}, on N1--4, respectively. We therefore find
the $B-$band to fade at a faster rate than the $V-$band, as expected,
and the slope in each filter to vary from night to night. The decrease
in the slope, observed in both bands, corresponds to the onset of the
secondary maximum in the near-IR light curve \citep{Kasen06, Jack15},
which is also seen in the light curves of SN~2014J presented by
\citet{Foley14}, \citet{Amanullah14}, \citet{Marion15},
\citet{Kawabata14}, and \citet{Ashall14}.

\begin{table*}%t2
\caption{Decline rates $\alpha$ (mag day$^{-1}$) and values of
  $\sigma_w$, $\sigma_r$, $\sigma_N$ (mag) based on SN--S1 of the $V$
  and $B-$band light curves of SN 2014J.}\label{table:decline}
\begin{tabular}{llccccccccc}
\hline\hline
Night & Filter & $\alpha_{SN-S1}$ & $\alpha_{SN-S2}$ & $\alpha_{SN-S3}$ & $\alpha_{SN-S4}$ & $\alpha_{SN-S5}$ & $\alpha_{SN_{TFA}}$ & $\sigma_w$ & $\sigma_r$ & $\sigma_N$ \\
\hline
   1 & $V$ & $0.06\pm0.01$ & $0.15\pm0.01$  & $0.17\pm0.02$ & $0.15\pm0.03$  & $0.16\pm0.02$ & $0.14\pm0.02$ & 0.0086 & 0.0045 & 0.0048 \\
   2 & $V$ & $0.09\pm0.01$ & $0.09\pm0.01$  & $0.10\pm0.02$ & $0.08\pm0.03$  & $0.07\pm0.03$ & $0.07\pm0.01$ & 0.0107 & 0.0050 & 0.0053 \\
   3 & $V$ & $0.10\pm0.01$ & $0.07\pm0.02$  & $0.04\pm0.03$ & ...   &   $0.08\pm0.03$ & $0.09\pm0.04$ & 0.0177 & 0.0052 & 0.0061 \\
   4 & $V$ & $0.03\pm0.01$ & $0.04\pm0.01$  & $0.05\pm0.02$ & $0.02\pm0.03$  & $0.03\pm0.02$ & $0.04\pm0.01$ & 0.0131 & 0.0046 & 0.0052 \\
\hline                                           
   1 & $B$ & $0.09\pm0.01$ & $0.19\pm0.01$  & $0.24\pm0.03$ & $0.21\pm0.03$  & $0.17\pm0.05$ & $0.22\pm0.05$ & 0.0104 & 0.0025 & 0.0031 \\
   2 & $B$ & $0.14\pm0.01$ & $0.12\pm0.01$  & $0.12\pm0.03$ & $0.10\pm0.03$  & $0.07\pm0.04$ & $0.09\pm0.01$ & 0.0074 & 0.0033 & 0.0036 \\
   3 & $B$ & $0.16\pm0.01$ & $0.13\pm0.02$  & $0.11\pm0.03$ & ...  & $0.20\pm0.04$ & $0.15\pm0.03$ & 0.0186 & 0.0036 & 0.0050 \\
   4 & $B$ & $0.07\pm0.01$ & $0.06\pm0.02$  & $0.11\pm0.03$ & $0.09\pm0.04$  & $0.12\pm0.04$ & $0.06\pm0.01$ & 0.0128 & 0.0059 & 0.0063 \\
\hline                                           
\end{tabular}
\end{table*}

More importantly, we report evidence for rapid variability in both the
$V$ and $B-$band light curves of SN~2014J.  We find that the level of
variability varies from night to night and is best traced by the SN--S1
curve, due to the accuracy in the photometric measurements achieved for
these bright stars. N3 and N4 exhibit the largest activity with
sinusoidal-like variations of amplitude up to 0.05 mag, while
variability on other nights is typically at the 0.02--0.03 mag level.  A
Fourier analysis of the SN-S1 curve did not yield a significant
periodicity.

While the precision of our measurements, based on the 1$\sigma$ error
bars resulting from the aperture photometry, is at the 1.4 mmag level
for SN--S1 (and at the 0.6 mmag level for the binned curve), we must
account for correlated sources of error (red noise) to quantify the
significance of the variability detection. Sources of correlated error
include the changing airmass, drifting of the stars across the CCD and
other instrumental parameters. We follow the procedure outlined by
\citet{Pont06}, which estimates the amount of correlated noise by
calculating the dispersion $\sigma_N$ from the binned residuals (after
subtracting a best fit model, in our case the decline rate) as a
function of $N$ points and finds the values for the white ($\sigma_w$)
and red ($\sigma_r$) noise that best fit the equation:
\begin{equation}
\sigma_N^2=\frac{\sigma_w^2}{N}+\sigma_r^2.
\end{equation}
The estimated values of $\sigma_w$, $\sigma_r$ and $\sigma_N$ for each
night and filter are shown in the last 3 columns of
Table~\ref{table:decline}, using the SN--S1 data and $N=30$ ($\sim1$
hour), which was selected to represent the longest timescale of the
detected variability. Note, that the values computed with
$N=10,\,20,\,40$ were very similar. If rapid variability is present,
then the $\sigma_N$ values are overestimates, as all points of the light
curves were used for the calculation. Given the derived $\sigma_N$
values (3--6 mmag), we find the 0.02-0.05 mag variations to be
statistically significant. At face value, the 0.05 mag variation in N3
(at HJD=2456707.46--2456707.48) has a $8.2\sigma$ significance in $V$
and $10\sigma$ in $B$, and the 0.04 mag variation in N4 (at
HJD=2456708.48$-$2456708.52) is a $7.7\sigma$ detection in $V$, and a
$6.3\sigma$ detection in $B$. In N2, the 0.02 mag variation at
HJD=2456706.32 is significant at the $3.8\sigma$ level in $V$ and
$5.6\sigma$ in $B$, while in N1, the 0.03 mag variation (at
HJD=2456705.47--2456705.51) is at the $6.2\sigma$ ($V$) and $9.7\sigma$
($B$) level. The timescale of sinusoidal-like variations ranges from
15--60 min. On N3, the $B-$band is found to precede the $V-$band by
$\sim10$ min. The $B-V$ color gradually increases within each night,
displaying an rms scatter of about 0.01 mag.

% Table 2 available electronically only
\onltab{
\begin{table*}%t2
\caption{Calibrated differential photometry of SN 2014J compared to S1
  in the $V$ and $B-$bands.}\label{table:photom}
\begin{tabular}{lccc}
\hline\hline
HJD--2450000 & Filter & SN--S1 & $\sigma_{SN-S1}$\\
\hline
   6705.30817 & $V$ & 1.231  &  0.001 \\
   6705.31330 & $V$ & 1.224  &  0.001 \\
   6705.31679 & $V$ & 1.212  &  0.001 \\
   6705.31766 & $V$ & 1.220  &  0.001 \\
   6705.31853 & $V$ & 1.219  &  0.001 \\
\hline
\end{tabular}
\end{table*}
}% end of onltab

\section{Discussion and Conclusions}

The high-cadence, high-precision photometry obtained with the 2.3m
Aristarchos telescope was crucial in providing evidence for rapid
variability in the optical light curve of SN~2014J. A cadence of 30 min,
such as that of the Kepler Mission, would not have been sufficient to
detect the sinusoidal-like variations nor any variability pattern seen
with the 2 min cadence of our observations. At face value, our
photometric precision of 3--6 mmag (based on the estimation of red
noise) or 2 mmag (based on artificial star tests on N3) for SN--S1
allowed for statistically significant detections of variability at the
level of 0.02--0.05 mag on all nights, peaking on N3, implying that the
phenomenon is common or perhaps even ubiquitous in SNe, unless it is
produced by a mechanism related exclusively to the onset of the
secondary maximum.

Theoretical models of SN light curves \citep[e.g.][]{Kasen06b, Sim13,
  Fink14} do not make predictions on such short time scales and
therefore cannot be compared with the photometry presented here. We
propose one or a combination of the following scenarios for the origin
of the variability: (i) clumping of the ejecta, possibly caused by
structures of intermediate mass elements in the outer layers
\citep{Hole10}, %as a
%non-uniform density and temperature profile is likely to cause brighter
%clumps to appear intermittently and alter the average flux observed,
(ii) interaction of the ejecta with circumstellar material, inferred by
\citet{Foley14}, (iii) asymmetry of the ejecta, as the explosion is not
expected to be spherically symmetric \citep[as inferred from
  spectropolarimetry, e.g. see review by][]{Wang08}, (iv) the onset of
the secondary maximum, which corresponds to a sudden decrease in the
flux mean opacity due to the transition from doubly to singly ionized
iron group elements \citep[see e.g.][]{Pinto00}.

Using the photospheric velocity $\sim$2 weeks after maximum
\citep[$-10,600$ km s$^{-1}$; estimated from Fig.\ 15 of][]{Foley14}, we
estimate the SN to have a radius of 104 AU on day 17. The 0.02--0.05 mag
variability corresponds to a fractional change of $\sim$2--5\% in
magnitude and flux, respectively, or a $\sim$1--2.5\% fractional change
in radius assuming constant luminosity and that the variation is due to
a change of radius, i.e.\ 1--2.6 AU. The observed fluctuation, however,
is an average variation of flux over the projected surface of the young
remnant; the 1--2.6 AU radius change gives an estimate of the surface
area fluctuating. \citet{Siverd15} reported relatively high-cadence
photometry of SN 2014J with the 4.2cm Kilodegree Extremely Little
Telescope North (KELT-N), thereby placing a $4.5\%$ (3$\sigma$) upper
limit on short timescale variations. The evidence for variability at the
level of $2-5\%$ presented in this work is therefore consistent with the
results of the measurement with KELT-N.

In conclusion, we strongly encourage the community to undertake
high-cadence and high-precision monitoring campaigns of future, nearby
and bright supernovae to confirm the presence of rapid variability,
determine whether it occurs in both SNe Ia and II light curves, and
differentiate between the scenarios for its origin. If variability is
due to asphericity in the explosion, it will provide evidence for
distinguishing the nature of SNe Ia progenitors
\citep[see][]{Livio11}. Furthermore, light curve variability, if shown
to be ubiquitous, is likely to contribute to the scatter in the mean
magnitude of SN Ia calibrators, which remains one of the largest factors
in the uncertainty of the Hubble constant \citep[e.g.][]{Riess11}. In
any case, we expect rapid variability to provide a new window into the
physics of supernovae.
 
\begin{acknowledgements}
The authors thank the anonymous referees for constructive comments that
helped clarify and improve the presentation of our results. The data are
available upon request. A.Z.B. acknowledges helpful discussions with
Saurabh Jha, Stephen Williams, Kris Stanek and Mercedes
L{\'o}pez-Morales during the preparation of this manuscript. The authors
acknowledge use of the IDL implementation of the red noise calculation
by J. Rogers and the excellent support of John Alikakos during the
observations. This research has made use of NASA's Astrophysics Data
System.
\end{acknowledgements}

%-------------------------------------------------------------------

\end{document}